\documentclass[
reprint,
 superscriptaddress,
 amsmath,amssymb,
 aps,
pra,
 floatfix,
]{revtex4-1}

\usepackage[T1]{fontenc} 
\usepackage{graphicx} 
\usepackage{dcolumn} 
\usepackage{bm} 
\usepackage[colorlinks,breaklinks,citecolor=blue,linkcolor=blue,urlcolor=blue]{hyperref}

\usepackage{xcolor}

\begin{document}
\title{A general Zeeman slower for type-II transitions and polar molecules}%

\author{Qian Liang}
\author{Wenhao Bu}
\author{Yuhe Zhang}
\author{Tao Chen}%
\email{phytch@zju.edu.cn}
\affiliation{%
Zhejiang Province Key Laboratory of Quantum Technology and Device, Department of Physics and State Key Laboratory of Modern Optical Instrumentation, Zhejiang University, Hangzhou, Zhejiang, 310027, China
}%
\author{Bo Yan}
\email{yanbohang@zju.edu.cn}
\affiliation{%
Zhejiang Province Key Laboratory of Quantum Technology and Device, Department of Physics and State Key Laboratory of Modern Optical Instrumentation, Zhejiang University, Hangzhou, Zhejiang, 310027, China
}%
\affiliation{%
Collaborative Innovation Centre of Advanced Microstructures, Nanjing University, Nanjing 210093, China
}%
\affiliation{%
Key Laboratory of Quantum Optics, Chinese Academy of Sciences, Shanghai 200800, China
}%

\begin{abstract}
We proposed a general Zeeman slower scheme applicable to the majority of the laser-coolable molecules. Different from previous schemes, the key idea of our scheme lies in that the compensation of the detuning with the magnetic field is done for the 
repumping laser instead of the cooling laser. Only atoms or molecules with the right velocity will be repumped and laser slowed.  Such scheme is more feasible for molecules with complex energy sturcutres. 
We apply this scheme for molecules with large Land\'e g-factor of the excited states and polyatomic molecules, and it shows a better slowing efficiency.
\end{abstract}
\maketitle

\section{Introduction}

Ultracold molecules provide new research opportunities on precision measurement \cite{Chin_2009, DeMille_2008, Safronova_2018}, ultracold chemistry \cite{Ospelkaus_2010}, many-body physics \cite{Wang_2006, Bloch_2008,Yan2013} and quantum computation 
and simulation \cite{DeMille_2002, Andre_2006, Micheli_2006, Rabl_2006}. While quantum ensembles below $1~\mu\text{K}$ have been achieved in some specific diatomic molecules by associating two ultracold alkali atoms 
\cite{Ni_2008,Bohn_2017}, direct laser cooling on diatomic and polyatomic molecules with highly diagonal Franck-Condon factors \cite{Barry_2014, Collopy_2018} recently has been successfully developed to produce cold molecular samples with 
a temperature below $50~\mu\text{K}$ \cite{Anderegg_2018, Np_Truppe_2017}. To push the direct cooling method further and achieve a higher phase-space density, a steady and efficient high-flux molecular source with slow enough velocity 
distribution that can be captured by a magneto-optical trap (MOT) is extremely required.

Although there are various molecular beam slowing schemes, such as the buffer gas cooling \cite{Maxwell_2005}, white-light slowing \cite{Barry_2012, Hemmerling_2016}, frequency-chirped slowing \cite{Truppe_2017, 
Yeo_2015, Zhelyazkova_2014}, optoelectric slowing and cooling \cite{Prehn_2016} and Stark/Zeeman deceleration \cite{Narevicius_2008,Meerakker_2006}, the poor controlling on the final velocity and the low 
slowing efficiency make the loading rate of MOT still limited. For example, the white-light slowing is continuous but the velocity distribution can not be compressed \cite{Zhu_1991, Barry_2012}. The widely used 
chirped slowing, however, can compress the velocity distribution in time but leads to a pulsed distribution of the slowed particles \cite{Ertmer_1985}. In practice, one should overcome these limitations to 
increase the number of molecules with lower enough velocities. The Zeeman slower combines advantages of the aforementioned schemes, including the continuous slowing, the compression of the velocity 
distribution and very good control on the final slowed velocity. It is promising to achieve a higher efficiency of molecular beam slowing and further a better implementation of MOT with the Zeeman slower. 

The traditional Zeeman slower working on atoms can't be directly applied to molecules due to their complex internal structures. Recently, a type-II Zeeman slower has been proposed \cite{Petzold_2018} 
and experimentally demonstrated for $^{39}$K atom with the D1-line transition \cite{Petzold_2018_pra}. The magnetic field applied is in a rather strong regime up to $\sim 1000~\text{G}$ to enter the Pachen-Back regime, so the energy gap between every two hyperfine states in the same $m_s$ manifold keeps almost the same under different magnetic field strength. In this way, the slowing scheme becomes greatly simplified and making the Zeeman slower feasible 
[see Fig.\ref{fig1}(a)]. The simulations in Ref.\cite{Petzold_2018} show that this scheme works well on the $|X^2\Sigma_{1/2}, N=1, v=0\rangle \to |A^2\Pi_{1/2}, J'=1/2, v'=0\rangle$ 
transition for SrF molecule where the Land\'e \emph{g}-factor of the excited state $g_\Pi$ is much smaller than that of the ground state $g_\Sigma$. However, it seems not so good for molecules with a large \emph{g}-factor for 
the upper state, for example, the BaF molecule with $g_{\Pi} \approx -0.202$. To go beyond this limitation, we come up with a different and more general Zeeman slowing scheme named dark state Zeeman slower (DSZS). Instead of 
compensating the Doppler shift with the magnetic field on the cooling laser, we make such compensations on the repump laser; see Fig.\ref{fig1}(b). Our simulations show that the slowing effect of our scheme 
is much better than that in Ref.\cite{Petzold_2018} in some cases. In the following, we will discuss how this dark state Zeeman slower scheme works in Sec.\ref{sec1}. Then, in Sec.\ref{sec2}, we numberically simulate the one-dimension deceleration 
process of the BaF molecule \cite{Chen_2016, Bu_2017, Chen2017} with the rate equation approach, and compare the slowing result with that from the aforementioned type-II Zeeman slower scheme. Furthermore, we extend the DSZS to a laser-coolable 
polyatomic molecule, SrOH \cite{Kozyryev_2015, Kozyryev_2016}, to check its applicability and universality.

\section{Scheme description and method}\label{sec1}

\begin{figure}[]
\includegraphics[width=0.48\textwidth]{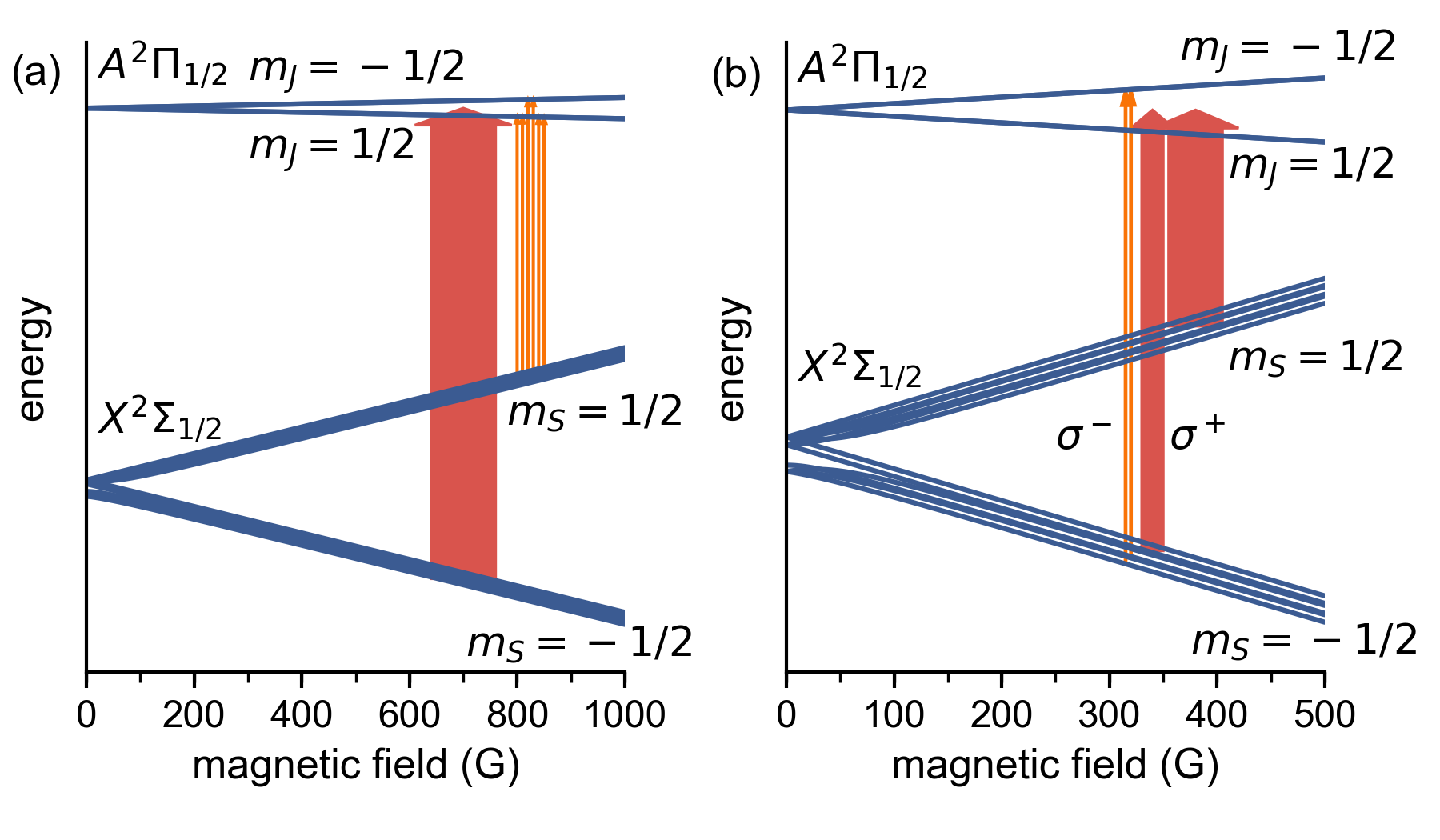}
\caption{\label{fig1} Two different Zeeman slowing schemes for diatomic molecules: (a) laser scheme from Ref.\cite{Petzold_2018} and (b) laser scheme in DSZS. The hyperfine levels of the ground state 
$X^2\Sigma_{1/2}$ are separated as two manifolds with $m_S = \pm1/2$ respectively. Each manifold has six sublevels. The excited state $A^2\Pi_{1/2}$ has four sublevels but the $m_I =\pm1/2$ states are 
degenerated since the nuclear spin g-factor is too small to affect the total Hamiltonian. For SrF molecule in (a), the splitting of the $m_J = \pm1/2$ states is negligible while it is not the case for BaF 
molecule in (b). The wide arrow (red) indicate the frequency-broadened repump laser and the narrow arrows indicate the frequency-modulated slowing laser in (a). In DSZS scheme (b), the wide red arrows are 
clean-up lasers while the narrow orange arrows are repump laser.}
\end{figure}

In a standard Zeeman slower \cite{Phillips_1982}, a position-dependent magnetic field is built to compensate the change of the Doppler shift as the particles are 
continually decelarated. The slowing laser always keeps on- or near-resonance with the main cooling transition to make the particles efficiently slowed during the propagation. 
Both the type-I and type-II Zeeman slowers share such a basic compensation mechanism. But for type-II ones, the particles may accumulate into the unwanted dark states, making the 
slowing force vanish. Even if we destabilize the dark states by applying an angled magnetic field \cite{Berkeland_2002}, the nonlinear energy shifts of the hyperfine sublevels for molecules make the conventional Zeeman slower no longer work well, especially 
in the weak magnetic field regime. In Ref.\cite{Petzold_2018}, a large offset magnetic field $B_0\sim 900~\text{G}$ is introduced to simplify the choice of the cooling and repumping laser components; 
see Fig.\ref{fig1}(a). Under a large magnetic field, the electron spin decouples with the nuclear spin and rotational angular momentum. The sublevels in the ground state now are separated 
into two manifolds with $m_S = \pm 1/2$ respectively. The excited state, however, can be regarded as degenerated due to the small \emph{g}-factor compared to the ground state [for example, 
SrF \cite{Shuman_2010} and YO \cite{Hummon_2013} molecules]. This makes the whole system reduce to a simple three-level system. Here, a slowing laser with six frequency sidebands couples the $m_S = +1/2$ states to the excited 
states according to the selection rules, while a frequency-broadened laser repumps the 
molecules that decay to the $m_S = -1/2$ manifold back to $m_S = +1/2$ for a wide velocity range. The repump laser must have a rather wide frequency broadening up to about $1~\text{GHz}$. In order to avoid the 
influence from the repump laser on the $m_S = -1/2$ manifold, the offset magnetic field should be large enough to split the $m_S = \pm 1/2$ manifolds away from each other. This type-II Zeeman 
slower has been proved to work as efficiently as the type-I Zeeman slower in a K atom experiment \cite{Petzold_2018_pra}.

It is obvious that the scheme only fits for molecules with a small upper \emph{g}-factor $g_{\Pi}$, otherwise the simplified three-level model collapses. When $g_{\Pi}$ is no longer small enough, the splitting 
between the upper $m_{J'} = \pm 1/2$ states is considerably large even under a small magnetic field. Let us revisit the scheme in Fig.\ref{fig1}(a), since the energy gap between the two upper $m_{J'}$ 
manifolds is large, the resonant frenquencies of the six cooling transitions can not change synchronously with the compensations from the magnetic field at the same time, leading to a less efficient slowing. 
Therefore, we propose the dark state Zeeman slower scheme (DSZS) shown in Fig.\ref{fig1}(b), in which the compensation of the Doppler shift is done on the repump laser. Similarly, we still apply an offset magnetic field to let the electron spin decoupled. Instead of applying a slowing 
laser with six frequencies to couple six hyperfine levels to the excited state, we use a repump laser with two frequencies to couple only two of the levels, which we call the \emph{'selected'} states. To ensure the molecules stay in the two \emph{selected} states, widely broadened clean-up beams are required to couple all the rest ground states at an arbitrary position during the propagation. In this way, it keeps molecules with the "wrong" velocity in the dark states while release them at the "right" 
velocity to get slowed at a specific position. When 
molecules with a velocity mismatch the resonant condition, they will be protected in the two \emph{selected} levels and wait for the Zeeman shift to match again. In the end, the velocity distribution of the molecular beam can 
be compressed into a sharp and narrow distribution around a tunable final velocity. 

To show the advantages of the DSZS scheme in Fig.\ref{fig1}(b), we take BaF molecule as an example. We choose the $|m_S = -1/2, m_N = +1, m_I = \pm 1/2\rangle$ states as the \emph{selected} states and the laser 
polarization to be $\sigma^-$ to make them only coupled to the excited $|m_{J'} = -1/2, m_{I'} = \pm 1/2\rangle$ states respectively. The polarization of the laser that drives all the
other states in the $m_S = -1/2$ manifold is chosen to be $\sigma^+$, such that the selected transitions will not be disturbed by it with a different polarization. 
Another clean-up laser that drives the ground $m_S = +1/2$ manifold is also required, and similarly we should apply an offset magnetic field to yield a large energy gap between the two ground manifolds. 
Because the repump laser and the clean-up lasers interact with the moving molecule independently, there is no need to keep a large energy gap between the $m_S=\pm1/2$ manifolds. So the DSZS can effectively work in a relatively smaller magnetic field regime, for example, $300~\text{G} 
\sim 400~\text{G}$ for the BaF molecule. In our scheme, the Land\'e g-factor of the excited state is no longer a restriction since the repump laser couples the \emph{selected} states to the same upper $m_{J'}$ 
manifold.

\begin{figure}[]
\includegraphics[width=0.45\textwidth]{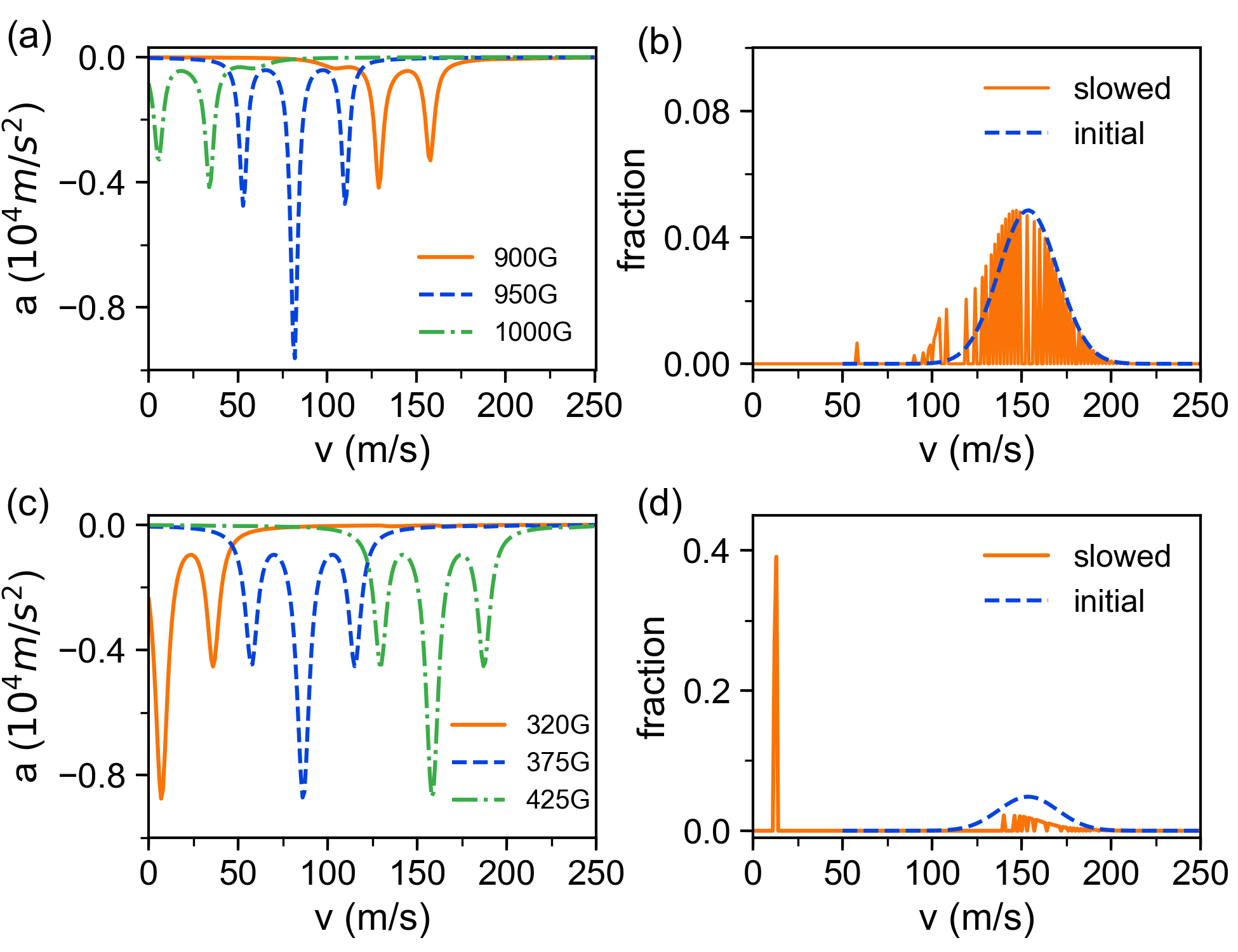}
\caption{\label{fig2}(a) The velocity-dependent deceleration rate under different magnetic field for the type-II Zeeman slower scheme in Fig.\ref{fig1}(a). 
(b) The initial velocity distribution (blue dotted line) and the final veloctiy distribution after the deceleration of the type-II Zeeman slower (orange solid line).
(c) The velocity dependent deceleration rate under different magnetic field with DSZS scheme shown in Fig.\ref{fig1}(b). 
(d) The initial velocity distribution (blue dotted line) and the final veloctiy distribution after the deceleration of DSZS (orange solid line).}
\end{figure}

We use the multi-level rate equation approach to calculate the slowing force for different moving velocities, and then simulate the slowing process of a molecular beam with an initial velocity distribution \cite{Truppe_2018}
\begin{equation}\label{eq1}
f(v) = Av^2 e^{-\beta (v-v_0)^2}
\end{equation}
where $A$ is the normalization parameter, $\beta$ is related to the temperature and the molecular mass, and $v_0$ is the center velocity.
The multi-level rate equations can be derived from the optical Bloch equation as \cite{Scully_1997}
\begin{eqnarray}
\frac{\mathrm{d}N_l}{dt} &=& \sum_{u,p}R_{l,u,p}(N_u - N_l) + \Gamma \sum_{u}r_{l,u}N_u, \nonumber\\
\frac{\mathrm{d}N_u}{dt} &=& \sum_{l,p}R_{l,u,p}(N_l - N_u) - \Gamma N_u,
\end{eqnarray}
where $N_u$ and $N_l$ are the populations of the excited state $|u\rangle$ and ground state $|l\rangle$ respectively, $r_{l,u}$ is the branching ratio of the dacay from the upper $|u\rangle $ state to the 
ground $|l\rangle$ state, and $\Gamma$ is the decay rate of the excited state. The excitation rate from $|l\rangle$ to $|u\rangle$ by the $p$-th beam is given by
\begin{equation}
R_{l,u,p} = \frac{\Gamma}{2} \frac{s_{p}}{1 + 4\Delta^2_{l,u}/\Gamma^2}
\end{equation}
with $s_{p}$ is the saturation parameter. $\Delta_{l,u}$ is the detuning between $|u\rangle$ and $|l\rangle$, including both the Zeeman shift and the Doppler shift. The radiative force can be obtained from the 
number of the scattered photons, i.e.,
\begin{equation}
F = \sum_{l,u,p}\hbar k_pR_{l,u,p}(N_u-N_l)
\end{equation}
with $k_p$ the wavevector of the $p$-th beam.

\section{simulation and comparison}\label{sec2}
\subsection{BaF molecule}\label{sec2.1}

The BaF molecule is a candidate for direct laser cooling and the molecular structures and branching ratios have already been investigated in detail \cite{Chen_2016}. In Fig.\ref{fig2}, we show the 
different slowing results for the BaF molecule with the two schemes mentioned in Fig.\ref{fig1}. For the one shown in Fig.\ref{fig1}(a), we set the position-dependent magnetic field with the formula \cite{Petzold_2018_pra}
\begin{equation}\label{eq5}
B = B_0 + \alpha L (1 - \sqrt{1-z/L}),
\end{equation}
in which $B_0$ is the offset magnetic field, $\alpha$ is the magnetic field gradient, $z$ is the position of the molecule and $L$ is the total length of the slower.
Here we set $L=1.75~\text{m}$, $B_0 = 900~\text{G}$, $\alpha = 57~\text{G/m}$, leading to a maximum magnetic field of $B_{\text{max}} = 1000~\text{G}$.
The slowing laser is sideband modulated carefully to be on-resonance with each slowing transition at $B = 950~\text{G}$ and the velocity $v = 80~\text{m/s}$. The saturation parameter for each 
slowing beam is $s=8$, while the repump laser takes $s=72$ with a frequency bandwidth of $1.2~\text{GHz}$.

For DSZS scheme in Fig.\ref{fig1}(b), the magnetic field has a monotonically decreasing form
\begin{equation}\label{eq6}
B = B_0 + \alpha L \sqrt{1-z/L}.
\end{equation}
The difference between the two magnetic field, i.e., Eq.(\ref{eq5}) and Eq.(\ref{eq6}), comes from the choices of different $m_S$ quantum numbers of the ground states that do the compensation.
The parameters we use here are $B_0 = 320~\text{G}$, $\alpha = 60~\text{G/m}$, and $L = 1.75~\text{m}$. The saturation parameters of two frequency components in the repump laser are $s=2$, 
and they are set to be on-resonance respectively with the transitions of $|m_S=-1/2, m_N=+1, m_I=\pm1/2\rangle \to |m_{J'}=-1/2, m_{I'}=\pm1/2\rangle$ at specific position and velocity. The bandwidth of the clean-up laser 
that pumps the rest states in $m_S = -1/2$ manifold is broadened to be $400~\text{MHz}$, and its saturation parameter $s=8$. For the laser that pumps the $m_S = +1/2$ manifold, the bandwidth is 
$750~\text{MHz}$ and the saturation parameter $s=72$. The initial velocity distribution centering at $150~\text{m/s}$ (from $50~\text{m/s}$ to $250~\text{m/s}$) is estimated by Eq.(\ref{eq1}) to be a 
good approximation to the experimental measurement.

\begin{figure}[]
\includegraphics[width=0.45\textwidth]{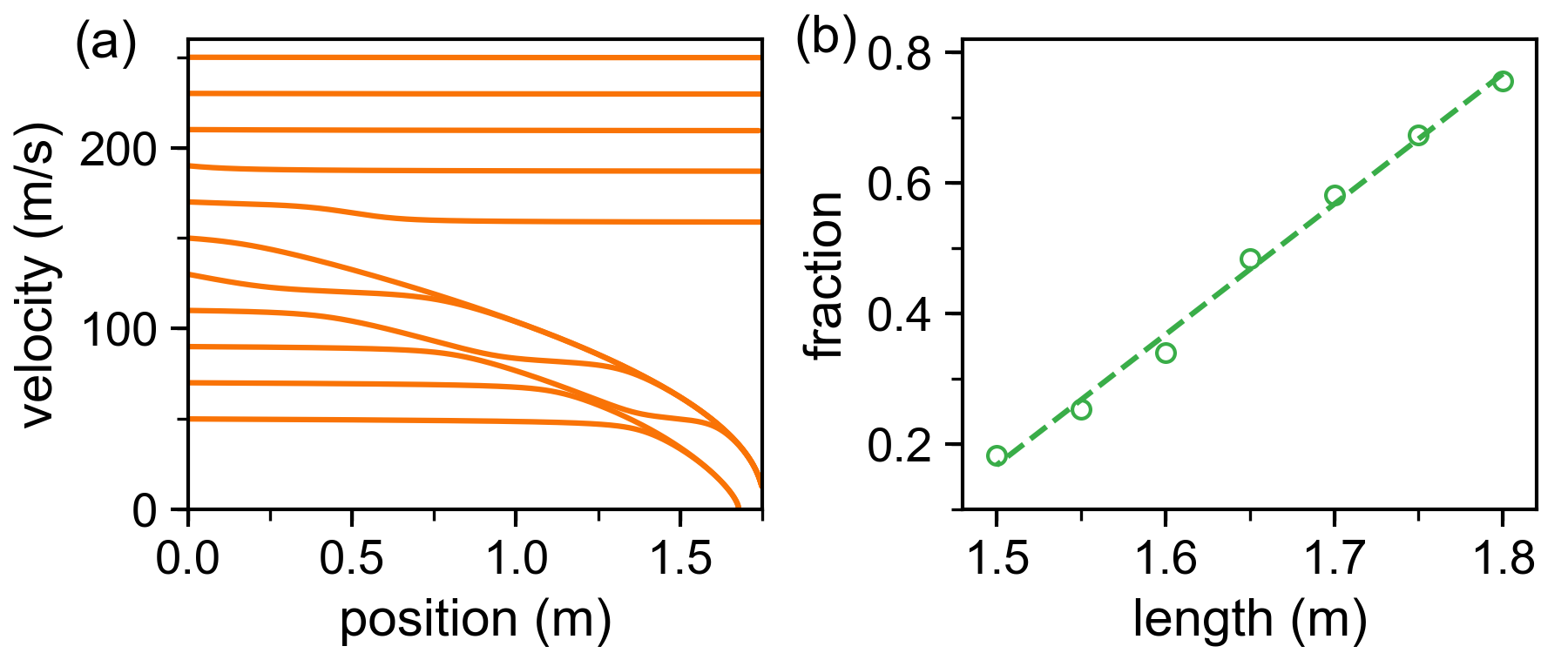}
\caption{\label{fig3}(a) The position-dependent evolutions of molecules with different initial velocities from $50~\text{m/s}$ to $250~\text{m/s}$.
(b) The dependence of the final compressed molecular fraction below $15~\text{m/s}$ on the length of the Zeeman slower.}
\end{figure}

As shown in Fig.\ref{fig2}(a), when the magnetic field changes, the type-II Zeeman slower scheme in Ref.\cite{Petzold_2018} can not provide a large and stable slowing force for the BaF molecule. As the magnetic field goes away from the 
center point $B = 950~\text{G}$, the absolute value of the deceleration rate rapidly decreases. The reason lies in that the states in $m_S = +1/2$ manifold are coupled to different excited 
states, i.e., $|m_S=+1/2, m_N=0, m_I=\pm1/2\rangle \to |m_{J'}=-1/2, m_{I'}=\pm1/2\rangle$ and $|m_S=+1/2, m_N=\pm1, m_I=\pm1/2\rangle \to |m_{J'}=+1/2, m_{I'}=\pm1/2\rangle$. The energy splittings of the 
$m_{J'} = \pm 1/2$ states show different behaviours when the magnetic field increases, resulting in that the slowing laser can not be resonant with the molecules all the time. 
 
The slowing force of our dark state Zeeman slower shows its robustness to the change of the magnetic field; see Fig.\ref{fig2}(c). By adjusting the magnetic field, a maximum deceleration rate of 
$-8000 ~\text{m/s}^2$ can be achieved. When all the frequency components of the repump beam are on-resonance respectively, no dark state exists and the deceleration rate reaches the maximum value, i.e., the 
center dip in Fig.\ref{fig2}(c). As the velocity changes, the repump beam becomes off resonance. The two \emph{selected} states become dark and the deceleration effect rapidly disappears. However, 
there is a special case that the Doppler shift only compensates for one frequency component to be resonant while makes the other far detuned. Then, one of the \emph{selected} states becomes dark and only the other one contributes to 
the photon scattering, resulting in the two weak dips besides the main center dip in Fig.\ref{fig2}(c).

A comparison of the slowing effect from the two different schemes is shown in Fig.\ref{fig2}. For the old one, with the parameters described above, the final velocity distribution slightly changes, 
and only a small fraction of BaF molecules are decelerated; see Fig.\ref{fig2}(b). However, with our DSZS scheme, the slowing effect gets enhanced. As shown in Fig.\ref{fig2}(d), about $66$\% of the BaF molecules 
are slowed to a velocity below $15~\text{m/s}$. Note that here we only consider one-dimensional slowing. In a real slowing experiment, the transverse divergence of the molecular beam should be taken 
into consideration \cite{Barry_2011}, which will weaken the low-velocity fraction of the molecules. Nevertheless, our simulations here show an effective slowing to make the beam converge to a 
narrow distribution around a low velocity. Such an enhancement makes it possible to load a MOT with a considerable molecular number.

Figure.\ref{fig3}(a) shows the trajectory of the moving molecules with different initial velocities when they pass through the dark state Zeeman slower. Molecules with a velocity larger than $\sim 170~\text{m/s}$ always 
stay in far-off detuning condition and no decelaration happens at all even as the magnetic field changes. For low-velocity molecules, for example, less than $100~\text{m/s}$, they are protected in the two 
\emph{selected} states at the early stage of the slowing, and then get slowed at a specific position where the Doppler shift and the Zeeman shift match with each other. The velocity range in which the 
molecules can be effectively slowed is determined by both the detuning of the repump laser and the distribution of the magnetic field (i.e., the gradient). The slowing efficiency also depends on the length of 
the slower. As shown in Fig.\ref{fig3}(b), the slowed fraction below 15 $\text{m/s}$ increases as the length $L$ increases. Here the offset magnetic field $B_0=320~\text{G}$ and the gradient 
$\alpha=60~\text{G/m}$ are fixed for various length in our simulation. Apparently, longer length leads to that molecules with higher velocity can be slowed down. Another issue is that the Zeeman slower for the 
BaF molecule should be long enough to get a better efficiency as the radiative force is small compared to some other laser-coolable molecules. In an actual experiment, a suitable compromise of the slower length 
and the transverse divergence should be taken into account.

\begin{figure}[b]
\includegraphics[width=0.3\textwidth]{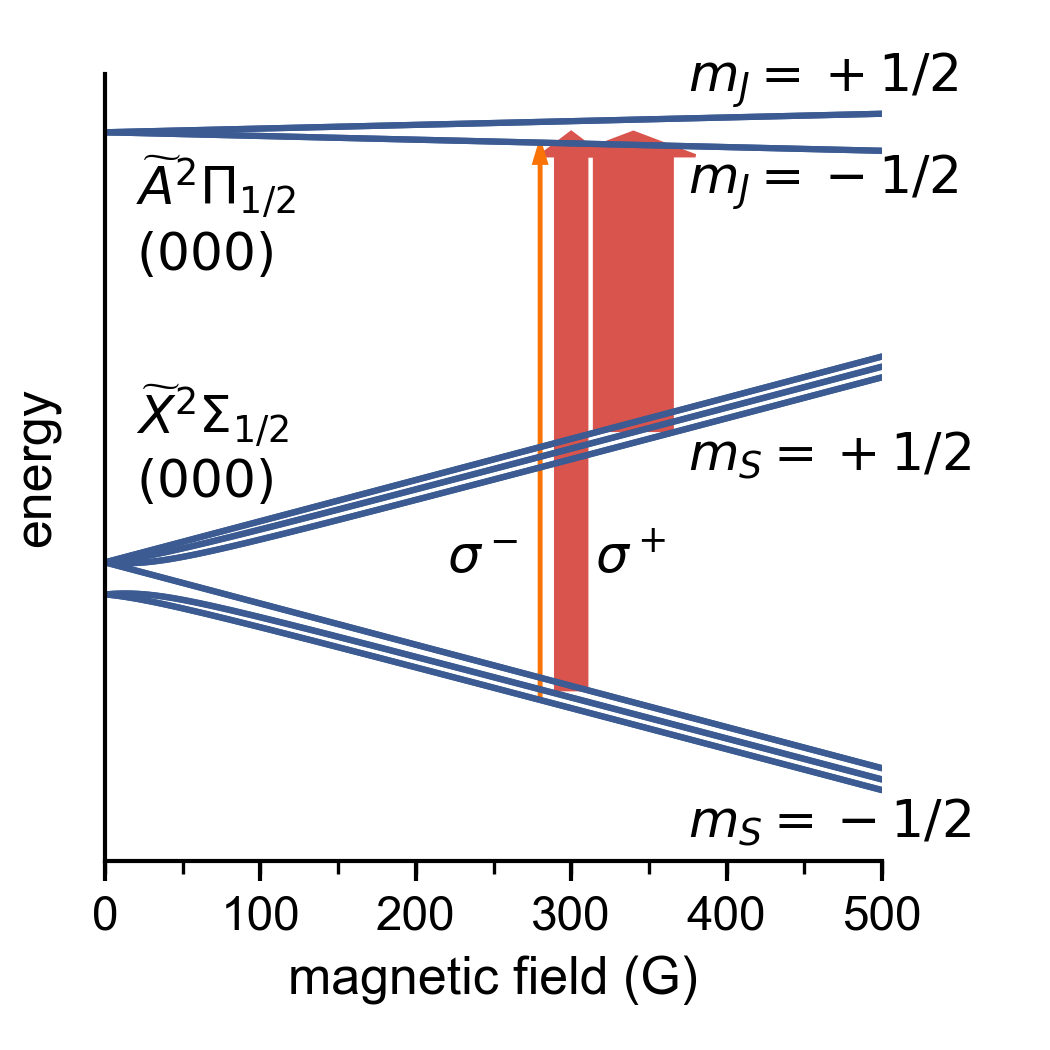}
\caption{\label{fig4}The structure of the $\widetilde{X}^2\Sigma_{1/2}(000)$ and $\widetilde{A}^2\Pi_{1/2}(000)$ of SrOH molecule (necessary data from Ref.\cite{Fletcher_1993}) 
 and the laser scheme applied.
The vibrational quantum numbers ($v_1v_2v_3$) correspond to the Sr$\leftrightarrow$OH stretching($v_1$) , Sr-O-H bending($v_2$) and SrO$\leftrightarrow$H stretching($v_3$)  vibrational modes. There are 12 
sublevels in $\widetilde{X}^2\Sigma_{1/2}(000)$ state but the $m_I = \pm1/2$ states are degenerated. A repump laser (the narrow orange arrow) is applied to couple the $|m_s = -1/2, m_N = 1, m_I = \pm1/2\rangle$ 
states with the corresponding excited states. Two clean-up lasers pumps the rest states (the wide red arrows)}
\end{figure}

\subsection{SrOH melocule}\label{sec2.2}

\begin{figure}[]
\includegraphics[width=0.5\textwidth]{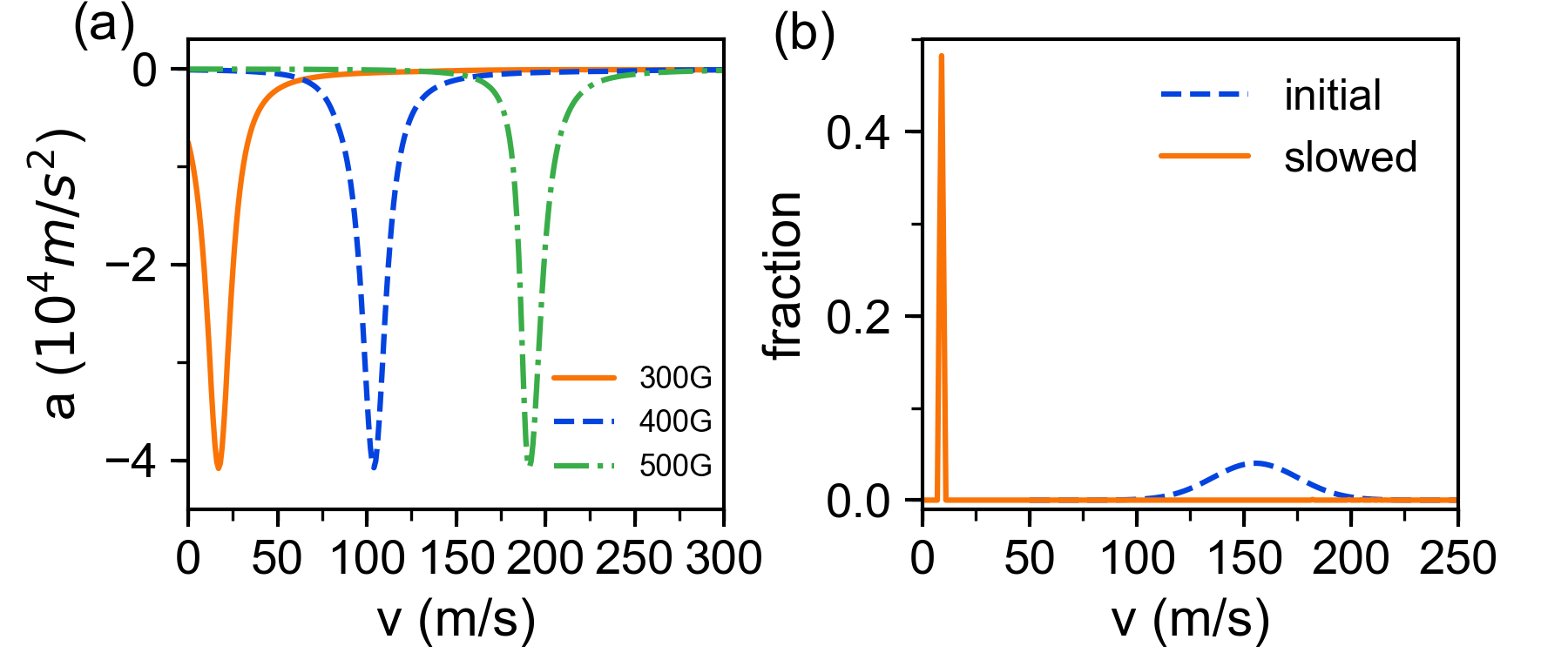}
\caption{\label{fig5} (a) The deceleration rate versus the velocity under different magnetic field.
(b) The initial velocity distribution (blue dotted line) and the final slowed veloctiy distribution (orange solid line). }
\end{figure}

To demonstrate the universality of our dark state Zeeman slower scheme, we check its applicability on a polyatomic molecule SrOH. The structure of the SrOH is much more complicated than the diatomic 
molecule for it contains three vibrational modes, including degenerated bending vibrations \cite{Presunka_1995}. Optical cycling on a quasi-closed transition and further sub-Doppler cooling have been realized 
\cite{Kozyryev_2016, Kozyryev_2017}. Here we work on the main cooling transition, i.e., $\widetilde{X}^2\Sigma_{1/2} \to \widetilde{A}^2\Pi_{1/2}$ transition. The splitting between the two 
$|m_S, m_N, m_I=\pm 1/2\rangle$ states is about $\sim 1~\text{MHz}$, and they are nearly degenerated, as shown in Fig.\ref{fig4}. Here only a single-frequency repump laser is enough, and no sideband 
modulations are in demand. In our simulation, the repump laser has a saturation paramter of $s=2$, coupling the ground states $|m_S=-1/2, m_N=1, m_I=\pm1/2\rangle$ to $|m_{J'}=-1/2, m_{I'}=\pm1/2\rangle$. 
The clean-up laser coupling the rest states in the $m_S=-1/2$ manifold has a saturation parameter of $s=8$ and a broadened frequency width of $180~\text{MHz}$, 
while another clean-up laser with $s=72$ and a broadened bandwidth of $700~\text{MHz}$. 

Figure.\ref{fig5}(a) shows the velocity-dependent deceleration rate for the SrOH molecule. The absolute value of deceleration rate can be up to $-4\times 10^4~\text{m/s}^2$, which is considerable large enough to 
slow the molecules. The slowing forces are stable and share a similar shape under different magnetic field strength. This indicates that our scheme can be directly extended to the SrOH molecule. Note that 
because the repump laser has only one frequency, the deceleration profile under a specific magnetic field only has one dip in Fig.\ref{fig5}(a), different from that for the BaF molecule in Fig.\ref{fig2}(c).

The distribution of the magnetic field of the slower still follows the formula of Eq.(\ref{eq6}), ranging from $540~\text{G}$ to $292~\text{G}$. The length of the slower is $L=1~\text{m}$. Figure.\ref{fig5}(b) 
shows the slowing effect. About $99\%$ of the molecules get slowed below $11~\text{m/s}$ without consideration on the transverse divergence. The results discussed above show that the DSZS scheme 
is capable of slowing polyatomic molecules that has a quasi-closed cycling transition. 

\section{Conclusion}
In summary, we propose a general dark state Zeeman slower that can decelerate most species of molecules that can be laser cooled. Compared with the scheme in Ref.\cite{Petzold_2018}, such a scheme make it 
possible to slow molecules that have a relatively large \emph{g}-factor of the excited state. Different from the traditional Zeeman slower used in atoms, an offset magnetic field is indispensable, 
and some clean-up lasers are required to avoid molecules accumulating in dark states. In our scheme, the requirements on the magnetic field and the frequency-broadened clean-up laser are accessible 
in the current experiments. We extend this scheme to the polyatomic molecule and obtain a good slowing effect with appropriate parameters. Our simulations show that our scheme can efficiently slow molecules 
to a small velocity under $15~\text{m/s}$ and significantly compress the distribution of the velocity. We expect this dark state Zeeman slower serving as a molecular source to increase the MOT loading rate 
and molecular number in the future laser-cooling experiments.

\begin{acknowledgments}
We acknowledge the support from Nation Natural Science Foundation of China under Grand No.91636104, the National Key Research and Development Program of China under Grand No.2018YFA0307200. Natural Science Foundation of Zhejiang province under Grant No. LZ18A040001, and the Fundamental Research Funds for the Central Universities. 
\end{acknowledgments}

\bibliographystyle{apsrev4-1}
\bibliography{zeeman_slower} 

\begin{thebibliography}{45}%
\makeatletter
\providecommand \@ifxundefined [1]{%
 \@ifx{#1\undefined}
}%
\providecommand \@ifnum [1]{%
 \ifnum #1\expandafter \@firstoftwo
 \else \expandafter \@secondoftwo
 \fi
}%
\providecommand \@ifx [1]{%
 \ifx #1\expandafter \@firstoftwo
 \else \expandafter \@secondoftwo
 \fi
}%
\providecommand \natexlab [1]{#1}%
\providecommand \enquote  [1]{``#1''}%
\providecommand \bibnamefont  [1]{#1}%
\providecommand \bibfnamefont [1]{#1}%
\providecommand \citenamefont [1]{#1}%
\providecommand \href@noop [0]{\@secondoftwo}%
\providecommand \href [0]{\begingroup \@sanitize@url \@href}%
\providecommand \@href[1]{\@@startlink{#1}\@@href}%
\providecommand \@@href[1]{\endgroup#1\@@endlink}%
\providecommand \@sanitize@url [0]{\catcode `\\12\catcode `\$12\catcode
  `\&12\catcode `\#12\catcode `\^12\catcode `\_12\catcode `\%12\relax}%
\providecommand \@@startlink[1]{}%
\providecommand \@@endlink[0]{}%
\providecommand \url  [0]{\begingroup\@sanitize@url \@url }%
\providecommand \@url [1]{\endgroup\@href {#1}{\urlprefix }}%
\providecommand \urlprefix  [0]{URL }%
\providecommand \Eprint [0]{\href }%
\providecommand \doibase [0]{http://dx.doi.org/}%
\providecommand \selectlanguage [0]{\@gobble}%
\providecommand \bibinfo  [0]{\@secondoftwo}%
\providecommand \bibfield  [0]{\@secondoftwo}%
\providecommand \translation [1]{[#1]}%
\providecommand \BibitemOpen [0]{}%
\providecommand \bibitemStop [0]{}%
\providecommand \bibitemNoStop [0]{.\EOS\space}%
\providecommand \EOS [0]{\spacefactor3000\relax}%
\providecommand \BibitemShut  [1]{\csname bibitem#1\endcsname}%
\let\auto@bib@innerbib\@empty
\bibitem [{\citenamefont {Chin}\ \emph {et~al.}(2009)\citenamefont {Chin},
  \citenamefont {Flambaum},\ and\ \citenamefont {Kozlov}}]{Chin_2009}%
  \BibitemOpen
  \bibfield  {author} {\bibinfo {author} {\bibfnamefont {C.}~\bibnamefont
  {Chin}}, \bibinfo {author} {\bibfnamefont {V.~V.}\ \bibnamefont {Flambaum}},
  \ and\ \bibinfo {author} {\bibfnamefont {M.~G.}\ \bibnamefont {Kozlov}},\
  }\href {\doibase 10.1088/1367-2630/11/5/055048} {\bibfield  {journal}
  {\bibinfo  {journal} {New J. Phys}\ }\textbf {\bibinfo {volume} {11}},\
  \bibinfo {pages} {055048} (\bibinfo {year} {2009})}\BibitemShut {NoStop}%
\bibitem [{\citenamefont {DeMille}\ \emph {et~al.}(2008)\citenamefont
  {DeMille}, \citenamefont {Cahn}, \citenamefont {Murphree}, \citenamefont
  {Rahmlow},\ and\ \citenamefont {Kozlov}}]{DeMille_2008}%
  \BibitemOpen
  \bibfield  {author} {\bibinfo {author} {\bibfnamefont {D.}~\bibnamefont
  {DeMille}}, \bibinfo {author} {\bibfnamefont {S.~B.}\ \bibnamefont {Cahn}},
  \bibinfo {author} {\bibfnamefont {D.}~\bibnamefont {Murphree}}, \bibinfo
  {author} {\bibfnamefont {D.~A.}\ \bibnamefont {Rahmlow}}, \ and\ \bibinfo
  {author} {\bibfnamefont {M.~G.}\ \bibnamefont {Kozlov}},\ }\href {\doibase
  10.1103/PhysRevLett.100.023003} {\bibfield  {journal} {\bibinfo  {journal}
  {Phys. Rev. Lett.}\ }\textbf {\bibinfo {volume} {100}},\ \bibinfo {pages}
  {023003} (\bibinfo {year} {2008})}\BibitemShut {NoStop}%
\bibitem [{\citenamefont {Safronova}\ \emph {et~al.}(2018)\citenamefont
  {Safronova}, \citenamefont {Budker}, \citenamefont {DeMille}, \citenamefont
  {Kimball}, \citenamefont {Derevianko},\ and\ \citenamefont
  {Clark}}]{Safronova_2018}%
  \BibitemOpen
  \bibfield  {author} {\bibinfo {author} {\bibfnamefont {M.~S.}\ \bibnamefont
  {Safronova}}, \bibinfo {author} {\bibfnamefont {D.}~\bibnamefont {Budker}},
  \bibinfo {author} {\bibfnamefont {D.}~\bibnamefont {DeMille}}, \bibinfo
  {author} {\bibfnamefont {D.~F.~J.}\ \bibnamefont {Kimball}}, \bibinfo
  {author} {\bibfnamefont {A.}~\bibnamefont {Derevianko}}, \ and\ \bibinfo
  {author} {\bibfnamefont {C.~W.}\ \bibnamefont {Clark}},\ }\href {\doibase
  10.1103/RevModPhys.90.025008} {\bibfield  {journal} {\bibinfo  {journal}
  {Rev. Mod. Phys.}\ }\textbf {\bibinfo {volume} {90}},\ \bibinfo {pages}
  {025008} (\bibinfo {year} {2018})}\BibitemShut {NoStop}%
\bibitem [{\citenamefont {Ospelkaus}\ \emph {et~al.}(2010)\citenamefont
  {Ospelkaus}, \citenamefont {Ni}, \citenamefont {Wang}, \citenamefont
  {de~Miranda}, \citenamefont {Neyenhuis}, \citenamefont {Qu{\'e}m{\'e}ner},
  \citenamefont {Julienne}, \citenamefont {Bohn}, \citenamefont {Jin},\ and\
  \citenamefont {Ye}}]{Ospelkaus_2010}%
  \BibitemOpen
  \bibfield  {author} {\bibinfo {author} {\bibfnamefont {S.}~\bibnamefont
  {Ospelkaus}}, \bibinfo {author} {\bibfnamefont {K.-K.}\ \bibnamefont {Ni}},
  \bibinfo {author} {\bibfnamefont {D.}~\bibnamefont {Wang}}, \bibinfo {author}
  {\bibfnamefont {M.~H.~G.}\ \bibnamefont {de~Miranda}}, \bibinfo {author}
  {\bibfnamefont {B.}~\bibnamefont {Neyenhuis}}, \bibinfo {author}
  {\bibfnamefont {G.}~\bibnamefont {Qu{\'e}m{\'e}ner}}, \bibinfo {author}
  {\bibfnamefont {P.~S.}\ \bibnamefont {Julienne}}, \bibinfo {author}
  {\bibfnamefont {J.~L.}\ \bibnamefont {Bohn}}, \bibinfo {author}
  {\bibfnamefont {D.~S.}\ \bibnamefont {Jin}}, \ and\ \bibinfo {author}
  {\bibfnamefont {J.}~\bibnamefont {Ye}},\ }\href {\doibase
  10.1126/science.1184121} {\bibfield  {journal} {\bibinfo  {journal}
  {Science}\ }\textbf {\bibinfo {volume} {327}},\ \bibinfo {pages} {853}
  (\bibinfo {year} {2010})}\BibitemShut {NoStop}%
\bibitem [{\citenamefont {Wang}\ \emph {et~al.}(2006)\citenamefont {Wang},
  \citenamefont {Lukin},\ and\ \citenamefont {Demler}}]{Wang_2006}%
  \BibitemOpen
  \bibfield  {author} {\bibinfo {author} {\bibfnamefont {D.-W.}\ \bibnamefont
  {Wang}}, \bibinfo {author} {\bibfnamefont {M.~D.}\ \bibnamefont {Lukin}}, \
  and\ \bibinfo {author} {\bibfnamefont {E.}~\bibnamefont {Demler}},\ }\href
  {\doibase 10.1103/PhysRevLett.97.180413} {\bibfield  {journal} {\bibinfo
  {journal} {Phys. Rev. Lett.}\ }\textbf {\bibinfo {volume} {97}},\ \bibinfo
  {pages} {180413} (\bibinfo {year} {2006})}\BibitemShut {NoStop}%
\bibitem [{\citenamefont {Bloch}\ \emph {et~al.}(2008)\citenamefont {Bloch},
  \citenamefont {Dalibard},\ and\ \citenamefont {Zwerger}}]{Bloch_2008}%
  \BibitemOpen
  \bibfield  {author} {\bibinfo {author} {\bibfnamefont {I.}~\bibnamefont
  {Bloch}}, \bibinfo {author} {\bibfnamefont {J.}~\bibnamefont {Dalibard}}, \
  and\ \bibinfo {author} {\bibfnamefont {W.}~\bibnamefont {Zwerger}},\ }\href
  {\doibase 10.1103/RevModPhys.80.885} {\bibfield  {journal} {\bibinfo
  {journal} {Rev. Mod. Phys.}\ }\textbf {\bibinfo {volume} {80}},\ \bibinfo
  {pages} {885} (\bibinfo {year} {2008})}\BibitemShut {NoStop}%
\bibitem [{\citenamefont {Yan}\ \emph {et~al.}(2013)\citenamefont {Yan},
  \citenamefont {Moses}, \citenamefont {Gadway}, \citenamefont {Covey},
  \citenamefont {Hazzard}, \citenamefont {Rey}, \citenamefont {Jin},\ and\
  \citenamefont {Ye}}]{Yan2013}%
  \BibitemOpen
  \bibfield  {author} {\bibinfo {author} {\bibfnamefont {B.}~\bibnamefont
  {Yan}}, \bibinfo {author} {\bibfnamefont {S.~A.}\ \bibnamefont {Moses}},
  \bibinfo {author} {\bibfnamefont {B.}~\bibnamefont {Gadway}}, \bibinfo
  {author} {\bibfnamefont {J.~P.}\ \bibnamefont {Covey}}, \bibinfo {author}
  {\bibfnamefont {K.~R.~A.}\ \bibnamefont {Hazzard}}, \bibinfo {author}
  {\bibfnamefont {A.~M.}\ \bibnamefont {Rey}}, \bibinfo {author} {\bibfnamefont
  {D.~S.}\ \bibnamefont {Jin}}, \ and\ \bibinfo {author} {\bibfnamefont
  {J.}~\bibnamefont {Ye}},\ }\href {\doibase 10.1038/nature12483} {\bibfield
  {journal} {\bibinfo  {journal} {Nature}\ }\textbf {\bibinfo {volume} {501}},\
  \bibinfo {pages} {521} (\bibinfo {year} {2013})}\BibitemShut {NoStop}%
\bibitem [{\citenamefont {DeMille}(2002)}]{DeMille_2002}%
  \BibitemOpen
  \bibfield  {author} {\bibinfo {author} {\bibfnamefont {D.}~\bibnamefont
  {DeMille}},\ }\href {\doibase 10.1103/PhysRevLett.88.067901} {\bibfield
  {journal} {\bibinfo  {journal} {Phys. Rev. Lett.}\ }\textbf {\bibinfo
  {volume} {88}},\ \bibinfo {pages} {067901} (\bibinfo {year}
  {2002})}\BibitemShut {NoStop}%
\bibitem [{\citenamefont {Andr{\'e}}\ \emph {et~al.}(2006)\citenamefont
  {Andr{\'e}}, \citenamefont {DeMille}, \citenamefont {Doyle}, \citenamefont
  {Lukin}, \citenamefont {Maxwell}, \citenamefont {Rabl}, \citenamefont
  {Schoelkopf},\ and\ \citenamefont {Zoller}}]{Andre_2006}%
  \BibitemOpen
  \bibfield  {author} {\bibinfo {author} {\bibfnamefont {A.}~\bibnamefont
  {Andr{\'e}}}, \bibinfo {author} {\bibfnamefont {D.}~\bibnamefont {DeMille}},
  \bibinfo {author} {\bibfnamefont {J.~M.}\ \bibnamefont {Doyle}}, \bibinfo
  {author} {\bibfnamefont {M.~D.}\ \bibnamefont {Lukin}}, \bibinfo {author}
  {\bibfnamefont {S.~E.}\ \bibnamefont {Maxwell}}, \bibinfo {author}
  {\bibfnamefont {P.}~\bibnamefont {Rabl}}, \bibinfo {author} {\bibfnamefont
  {R.~J.}\ \bibnamefont {Schoelkopf}}, \ and\ \bibinfo {author} {\bibfnamefont
  {P.}~\bibnamefont {Zoller}},\ }\href {\doibase 10.1038/nphys386} {\bibfield
  {journal} {\bibinfo  {journal} {Nat. Phys}\ }\textbf {\bibinfo {volume}
  {2}},\ \bibinfo {pages} {636} (\bibinfo {year} {2006})}\BibitemShut {NoStop}%
\bibitem [{\citenamefont {Micheli}\ \emph {et~al.}(2006)\citenamefont
  {Micheli}, \citenamefont {Brennen},\ and\ \citenamefont
  {Zoller}}]{Micheli_2006}%
  \BibitemOpen
  \bibfield  {author} {\bibinfo {author} {\bibfnamefont {A.}~\bibnamefont
  {Micheli}}, \bibinfo {author} {\bibfnamefont {G.~K.}\ \bibnamefont
  {Brennen}}, \ and\ \bibinfo {author} {\bibfnamefont {P.}~\bibnamefont
  {Zoller}},\ }\href {\doibase 10.1038/nphys287} {\bibfield  {journal}
  {\bibinfo  {journal} {Nat. Phys}\ }\textbf {\bibinfo {volume} {2}},\ \bibinfo
  {pages} {341} (\bibinfo {year} {2006})}\BibitemShut {NoStop}%
\bibitem [{\citenamefont {Rabl}\ \emph {et~al.}(2006)\citenamefont {Rabl},
  \citenamefont {DeMille}, \citenamefont {Doyle}, \citenamefont {Lukin},
  \citenamefont {Schoelkopf},\ and\ \citenamefont {Zoller}}]{Rabl_2006}%
  \BibitemOpen
  \bibfield  {author} {\bibinfo {author} {\bibfnamefont {P.}~\bibnamefont
  {Rabl}}, \bibinfo {author} {\bibfnamefont {D.}~\bibnamefont {DeMille}},
  \bibinfo {author} {\bibfnamefont {J.~M.}\ \bibnamefont {Doyle}}, \bibinfo
  {author} {\bibfnamefont {M.~D.}\ \bibnamefont {Lukin}}, \bibinfo {author}
  {\bibfnamefont {R.~J.}\ \bibnamefont {Schoelkopf}}, \ and\ \bibinfo {author}
  {\bibfnamefont {P.}~\bibnamefont {Zoller}},\ }\href {\doibase
  10.1103/PhysRevLett.97.033003} {\bibfield  {journal} {\bibinfo  {journal}
  {Phys. Rev. Lett.}\ }\textbf {\bibinfo {volume} {97}},\ \bibinfo {pages}
  {033003} (\bibinfo {year} {2006})}\BibitemShut {NoStop}%
\bibitem [{\citenamefont {Ni}\ \emph {et~al.}(2008)\citenamefont {Ni},
  \citenamefont {Ospelkaus}, \citenamefont {de~Miranda}, \citenamefont
  {Pe{\textquoteright}er}, \citenamefont {Neyenhuis}, \citenamefont {Zirbel},
  \citenamefont {Kotochigova}, \citenamefont {Julienne}, \citenamefont {Jin},\
  and\ \citenamefont {Ye}}]{Ni_2008}%
  \BibitemOpen
  \bibfield  {author} {\bibinfo {author} {\bibfnamefont {K.-K.}\ \bibnamefont
  {Ni}}, \bibinfo {author} {\bibfnamefont {S.}~\bibnamefont {Ospelkaus}},
  \bibinfo {author} {\bibfnamefont {M.~H.~G.}\ \bibnamefont {de~Miranda}},
  \bibinfo {author} {\bibfnamefont {A.}~\bibnamefont {Pe{\textquoteright}er}},
  \bibinfo {author} {\bibfnamefont {B.}~\bibnamefont {Neyenhuis}}, \bibinfo
  {author} {\bibfnamefont {J.~J.}\ \bibnamefont {Zirbel}}, \bibinfo {author}
  {\bibfnamefont {S.}~\bibnamefont {Kotochigova}}, \bibinfo {author}
  {\bibfnamefont {P.~S.}\ \bibnamefont {Julienne}}, \bibinfo {author}
  {\bibfnamefont {D.~S.}\ \bibnamefont {Jin}}, \ and\ \bibinfo {author}
  {\bibfnamefont {J.}~\bibnamefont {Ye}},\ }\href {\doibase
  10.1126/science.1163861} {\bibfield  {journal} {\bibinfo  {journal}
  {Science}\ }\textbf {\bibinfo {volume} {322}},\ \bibinfo {pages} {231}
  (\bibinfo {year} {2008})}\BibitemShut {NoStop}%
\bibitem [{\citenamefont {Bohn}\ \emph {et~al.}(2017)\citenamefont {Bohn},
  \citenamefont {Rey},\ and\ \citenamefont {Ye}}]{Bohn_2017}%
  \BibitemOpen
  \bibfield  {author} {\bibinfo {author} {\bibfnamefont {J.~L.}\ \bibnamefont
  {Bohn}}, \bibinfo {author} {\bibfnamefont {A.~M.}\ \bibnamefont {Rey}}, \
  and\ \bibinfo {author} {\bibfnamefont {J.}~\bibnamefont {Ye}},\ }\href
  {\doibase 10.1126/science.aam6299} {\bibfield  {journal} {\bibinfo  {journal}
  {Science}\ }\textbf {\bibinfo {volume} {357}},\ \bibinfo {pages} {1002}
  (\bibinfo {year} {2017})}\BibitemShut {NoStop}%
\bibitem [{\citenamefont {Barry}\ \emph {et~al.}(2014)\citenamefont {Barry},
  \citenamefont {McCarron}, \citenamefont {Norrgard}, \citenamefont
  {Steinecker},\ and\ \citenamefont {DeMille}}]{Barry_2014}%
  \BibitemOpen
  \bibfield  {author} {\bibinfo {author} {\bibfnamefont {J.~F.}\ \bibnamefont
  {Barry}}, \bibinfo {author} {\bibfnamefont {D.~J.}\ \bibnamefont {McCarron}},
  \bibinfo {author} {\bibfnamefont {E.~B.}\ \bibnamefont {Norrgard}}, \bibinfo
  {author} {\bibfnamefont {M.~H.}\ \bibnamefont {Steinecker}}, \ and\ \bibinfo
  {author} {\bibfnamefont {D.}~\bibnamefont {DeMille}},\ }\href {\doibase
  10.1038/nature13634} {\bibfield  {journal} {\bibinfo  {journal} {Nature}\
  }\textbf {\bibinfo {volume} {{512}}},\ \bibinfo {pages} {{286}} (\bibinfo
  {year} {{2014}})}\BibitemShut {NoStop}%
\bibitem [{\citenamefont {Collopy}\ \emph {et~al.}(2018)\citenamefont
  {Collopy}, \citenamefont {Ding}, \citenamefont {Wu}, \citenamefont
  {Finneran}, \citenamefont {Anderegg}, \citenamefont {Augenbraun},
  \citenamefont {Doyle},\ and\ \citenamefont {Ye}}]{Collopy_2018}%
  \BibitemOpen
  \bibfield  {author} {\bibinfo {author} {\bibfnamefont {A.~L.}\ \bibnamefont
  {Collopy}}, \bibinfo {author} {\bibfnamefont {S.}~\bibnamefont {Ding}},
  \bibinfo {author} {\bibfnamefont {Y.}~\bibnamefont {Wu}}, \bibinfo {author}
  {\bibfnamefont {I.~A.}\ \bibnamefont {Finneran}}, \bibinfo {author}
  {\bibfnamefont {L.}~\bibnamefont {Anderegg}}, \bibinfo {author}
  {\bibfnamefont {B.~L.}\ \bibnamefont {Augenbraun}}, \bibinfo {author}
  {\bibfnamefont {J.~M.}\ \bibnamefont {Doyle}}, \ and\ \bibinfo {author}
  {\bibfnamefont {J.}~\bibnamefont {Ye}},\ }\href {\doibase
  10.1103/PhysRevLett.121.213201} {\bibfield  {journal} {\bibinfo  {journal}
  {Phys. Rev. Lett.}\ }\textbf {\bibinfo {volume} {121}},\ \bibinfo {pages}
  {213201} (\bibinfo {year} {2018})}\BibitemShut {NoStop}%
\bibitem [{\citenamefont {Anderegg}\ \emph {et~al.}(2018)\citenamefont
  {Anderegg}, \citenamefont {Augenbraun}, \citenamefont {Bao}, \citenamefont
  {Burchesky}, \citenamefont {Cheuk}, \citenamefont {Ketterle},\ and\
  \citenamefont {Doyle}}]{Anderegg_2018}%
  \BibitemOpen
  \bibfield  {author} {\bibinfo {author} {\bibfnamefont {L.}~\bibnamefont
  {Anderegg}}, \bibinfo {author} {\bibfnamefont {B.~L.}\ \bibnamefont
  {Augenbraun}}, \bibinfo {author} {\bibfnamefont {Y.}~\bibnamefont {Bao}},
  \bibinfo {author} {\bibfnamefont {S.}~\bibnamefont {Burchesky}}, \bibinfo
  {author} {\bibfnamefont {L.~W.}\ \bibnamefont {Cheuk}}, \bibinfo {author}
  {\bibfnamefont {W.}~\bibnamefont {Ketterle}}, \ and\ \bibinfo {author}
  {\bibfnamefont {J.~M.}\ \bibnamefont {Doyle}},\ }\href {\doibase
  10.1038/s41567-018-0191-z} {\bibfield  {journal} {\bibinfo  {journal} {Nat.
  Phys}\ }\textbf {\bibinfo {volume} {14}},\ \bibinfo {pages} {890} (\bibinfo
  {year} {2018})}\BibitemShut {NoStop}%
\bibitem [{\citenamefont {Truppe}\ \emph
  {et~al.}(2017{\natexlab{a}})\citenamefont {Truppe}, \citenamefont {Williams},
  \citenamefont {Hambach}, \citenamefont {Caldwell}, \citenamefont {Fitch},
  \citenamefont {Hinds}, \citenamefont {Sauer},\ and\ \citenamefont
  {Tarbutt}}]{Np_Truppe_2017}%
  \BibitemOpen
  \bibfield  {author} {\bibinfo {author} {\bibfnamefont {S.}~\bibnamefont
  {Truppe}}, \bibinfo {author} {\bibfnamefont {H.~J.}\ \bibnamefont
  {Williams}}, \bibinfo {author} {\bibfnamefont {M.}~\bibnamefont {Hambach}},
  \bibinfo {author} {\bibfnamefont {L.}~\bibnamefont {Caldwell}}, \bibinfo
  {author} {\bibfnamefont {N.~J.}\ \bibnamefont {Fitch}}, \bibinfo {author}
  {\bibfnamefont {E.~A.}\ \bibnamefont {Hinds}}, \bibinfo {author}
  {\bibfnamefont {B.~E.}\ \bibnamefont {Sauer}}, \ and\ \bibinfo {author}
  {\bibfnamefont {M.~R.}\ \bibnamefont {Tarbutt}},\ }\href {\doibase
  10.1038/nphys4241} {\bibfield  {journal} {\bibinfo  {journal} {Nat. Phys}\
  }\textbf {\bibinfo {volume} {13}},\ \bibinfo {pages} {1173} (\bibinfo {year}
  {2017}{\natexlab{a}})}\BibitemShut {NoStop}%
\bibitem [{\citenamefont {Maxwell}\ \emph {et~al.}(2005)\citenamefont
  {Maxwell}, \citenamefont {Brahms}, \citenamefont {deCarvalho}, \citenamefont
  {Glenn}, \citenamefont {Helton}, \citenamefont {Nguyen}, \citenamefont
  {Patterson}, \citenamefont {Petricka}, \citenamefont {DeMille},\ and\
  \citenamefont {Doyle}}]{Maxwell_2005}%
  \BibitemOpen
  \bibfield  {author} {\bibinfo {author} {\bibfnamefont {S.~E.}\ \bibnamefont
  {Maxwell}}, \bibinfo {author} {\bibfnamefont {N.}~\bibnamefont {Brahms}},
  \bibinfo {author} {\bibfnamefont {R.}~\bibnamefont {deCarvalho}}, \bibinfo
  {author} {\bibfnamefont {D.~R.}\ \bibnamefont {Glenn}}, \bibinfo {author}
  {\bibfnamefont {J.~S.}\ \bibnamefont {Helton}}, \bibinfo {author}
  {\bibfnamefont {S.~V.}\ \bibnamefont {Nguyen}}, \bibinfo {author}
  {\bibfnamefont {D.}~\bibnamefont {Patterson}}, \bibinfo {author}
  {\bibfnamefont {J.}~\bibnamefont {Petricka}}, \bibinfo {author}
  {\bibfnamefont {D.}~\bibnamefont {DeMille}}, \ and\ \bibinfo {author}
  {\bibfnamefont {J.~M.}\ \bibnamefont {Doyle}},\ }\href {\doibase
  10.1103/PhysRevLett.95.173201} {\bibfield  {journal} {\bibinfo  {journal}
  {Phys. Rev. Lett.}\ }\textbf {\bibinfo {volume} {95}},\ \bibinfo {pages}
  {173201} (\bibinfo {year} {2005})}\BibitemShut {NoStop}%
\bibitem [{\citenamefont {Barry}\ \emph {et~al.}(2012)\citenamefont {Barry},
  \citenamefont {Shuman}, \citenamefont {Norrgard},\ and\ \citenamefont
  {DeMille}}]{Barry_2012}%
  \BibitemOpen
  \bibfield  {author} {\bibinfo {author} {\bibfnamefont {J.~F.}\ \bibnamefont
  {Barry}}, \bibinfo {author} {\bibfnamefont {E.~S.}\ \bibnamefont {Shuman}},
  \bibinfo {author} {\bibfnamefont {E.~B.}\ \bibnamefont {Norrgard}}, \ and\
  \bibinfo {author} {\bibfnamefont {D.}~\bibnamefont {DeMille}},\ }\href
  {\doibase 10.1103/PhysRevLett.108.103002} {\bibfield  {journal} {\bibinfo
  {journal} {Phys. Rev. Lett.}\ }\textbf {\bibinfo {volume} {108}},\ \bibinfo
  {pages} {103002} (\bibinfo {year} {2012})}\BibitemShut {NoStop}%
\bibitem [{\citenamefont {Hemmerling}\ \emph {et~al.}(2016)\citenamefont
  {Hemmerling}, \citenamefont {Chae}, \citenamefont {Ravi}, \citenamefont
  {Anderegg}, \citenamefont {Drayna}, \citenamefont {Hutzler}, \citenamefont
  {Collopy}, \citenamefont {Ye}, \citenamefont {Ketterle},\ and\ \citenamefont
  {Doyle}}]{Hemmerling_2016}%
  \BibitemOpen
  \bibfield  {author} {\bibinfo {author} {\bibfnamefont {B.}~\bibnamefont
  {Hemmerling}}, \bibinfo {author} {\bibfnamefont {E.}~\bibnamefont {Chae}},
  \bibinfo {author} {\bibfnamefont {A.}~\bibnamefont {Ravi}}, \bibinfo {author}
  {\bibfnamefont {L.}~\bibnamefont {Anderegg}}, \bibinfo {author}
  {\bibfnamefont {G.~K.}\ \bibnamefont {Drayna}}, \bibinfo {author}
  {\bibfnamefont {N.~R.}\ \bibnamefont {Hutzler}}, \bibinfo {author}
  {\bibfnamefont {A.~L.}\ \bibnamefont {Collopy}}, \bibinfo {author}
  {\bibfnamefont {J.}~\bibnamefont {Ye}}, \bibinfo {author} {\bibfnamefont
  {W.}~\bibnamefont {Ketterle}}, \ and\ \bibinfo {author} {\bibfnamefont
  {J.~M.}\ \bibnamefont {Doyle}},\ }\href {\doibase
  10.1088/0953-4075/49/17/174001} {\bibfield  {journal} {\bibinfo  {journal}
  {J. Phys. B}\ }\textbf {\bibinfo {volume} {49}},\ \bibinfo {pages} {174001}
  (\bibinfo {year} {2016})}\BibitemShut {NoStop}%
\bibitem [{\citenamefont {Truppe}\ \emph
  {et~al.}(2017{\natexlab{b}})\citenamefont {Truppe}, \citenamefont {Williams},
  \citenamefont {Fitch}, \citenamefont {Hambach}, \citenamefont {Wall},
  \citenamefont {Hinds}, \citenamefont {Sauer},\ and\ \citenamefont
  {Tarbutt}}]{Truppe_2017}%
  \BibitemOpen
  \bibfield  {author} {\bibinfo {author} {\bibfnamefont {S.}~\bibnamefont
  {Truppe}}, \bibinfo {author} {\bibfnamefont {H.~J.}\ \bibnamefont
  {Williams}}, \bibinfo {author} {\bibfnamefont {N.~J.}\ \bibnamefont {Fitch}},
  \bibinfo {author} {\bibfnamefont {M.}~\bibnamefont {Hambach}}, \bibinfo
  {author} {\bibfnamefont {T.~E.}\ \bibnamefont {Wall}}, \bibinfo {author}
  {\bibfnamefont {E.~A.}\ \bibnamefont {Hinds}}, \bibinfo {author}
  {\bibfnamefont {B.~E.}\ \bibnamefont {Sauer}}, \ and\ \bibinfo {author}
  {\bibfnamefont {M.~R.}\ \bibnamefont {Tarbutt}},\ }\href {\doibase
  10.1088/1367-2630/aa5ca2} {\bibfield  {journal} {\bibinfo  {journal} {New J.
  Phys}\ }\textbf {\bibinfo {volume} {19}},\ \bibinfo {pages} {022001}
  (\bibinfo {year} {2017}{\natexlab{b}})}\BibitemShut {NoStop}%
\bibitem [{\citenamefont {Yeo}\ \emph {et~al.}(2015)\citenamefont {Yeo},
  \citenamefont {Hummon}, \citenamefont {Collopy}, \citenamefont {Yan},
  \citenamefont {Hemmerling}, \citenamefont {Chae}, \citenamefont {Doyle},\
  and\ \citenamefont {Ye}}]{Yeo_2015}%
  \BibitemOpen
  \bibfield  {author} {\bibinfo {author} {\bibfnamefont {M.}~\bibnamefont
  {Yeo}}, \bibinfo {author} {\bibfnamefont {M.~T.}\ \bibnamefont {Hummon}},
  \bibinfo {author} {\bibfnamefont {A.~L.}\ \bibnamefont {Collopy}}, \bibinfo
  {author} {\bibfnamefont {B.}~\bibnamefont {Yan}}, \bibinfo {author}
  {\bibfnamefont {B.}~\bibnamefont {Hemmerling}}, \bibinfo {author}
  {\bibfnamefont {E.}~\bibnamefont {Chae}}, \bibinfo {author} {\bibfnamefont
  {J.~M.}\ \bibnamefont {Doyle}}, \ and\ \bibinfo {author} {\bibfnamefont
  {J.}~\bibnamefont {Ye}},\ }\href {\doibase 10.1103/PhysRevLett.114.223003}
  {\bibfield  {journal} {\bibinfo  {journal} {Phys. Rev. Lett.}\ }\textbf
  {\bibinfo {volume} {114}},\ \bibinfo {pages} {223003} (\bibinfo {year}
  {2015})}\BibitemShut {NoStop}%
\bibitem [{\citenamefont {Zhelyazkova}\ \emph {et~al.}(2014)\citenamefont
  {Zhelyazkova}, \citenamefont {Cournol}, \citenamefont {Wall}, \citenamefont
  {Matsushima}, \citenamefont {Hudson}, \citenamefont {Hinds}, \citenamefont
  {Tarbutt},\ and\ \citenamefont {Sauer}}]{Zhelyazkova_2014}%
  \BibitemOpen
  \bibfield  {author} {\bibinfo {author} {\bibfnamefont {V.}~\bibnamefont
  {Zhelyazkova}}, \bibinfo {author} {\bibfnamefont {A.}~\bibnamefont
  {Cournol}}, \bibinfo {author} {\bibfnamefont {T.~E.}\ \bibnamefont {Wall}},
  \bibinfo {author} {\bibfnamefont {A.}~\bibnamefont {Matsushima}}, \bibinfo
  {author} {\bibfnamefont {J.~J.}\ \bibnamefont {Hudson}}, \bibinfo {author}
  {\bibfnamefont {E.~A.}\ \bibnamefont {Hinds}}, \bibinfo {author}
  {\bibfnamefont {M.~R.}\ \bibnamefont {Tarbutt}}, \ and\ \bibinfo {author}
  {\bibfnamefont {B.~E.}\ \bibnamefont {Sauer}},\ }\href {\doibase
  10.1103/PhysRevA.89.053416} {\bibfield  {journal} {\bibinfo  {journal} {Phys.
  Rev. A}\ }\textbf {\bibinfo {volume} {89}},\ \bibinfo {pages} {053416}
  (\bibinfo {year} {2014})}\BibitemShut {NoStop}%
\bibitem [{\citenamefont {Prehn}\ \emph {et~al.}(2016)\citenamefont {Prehn},
  \citenamefont {Ibr\"ugger}, \citenamefont {Gl\"ockner}, \citenamefont
  {Rempe},\ and\ \citenamefont {Zeppenfeld}}]{Prehn_2016}%
  \BibitemOpen
  \bibfield  {author} {\bibinfo {author} {\bibfnamefont {A.}~\bibnamefont
  {Prehn}}, \bibinfo {author} {\bibfnamefont {M.}~\bibnamefont {Ibr\"ugger}},
  \bibinfo {author} {\bibfnamefont {R.}~\bibnamefont {Gl\"ockner}}, \bibinfo
  {author} {\bibfnamefont {G.}~\bibnamefont {Rempe}}, \ and\ \bibinfo {author}
  {\bibfnamefont {M.}~\bibnamefont {Zeppenfeld}},\ }\href {\doibase
  10.1103/PhysRevLett.116.063005} {\bibfield  {journal} {\bibinfo  {journal}
  {Phys. Rev. Lett.}\ }\textbf {\bibinfo {volume} {116}},\ \bibinfo {pages}
  {063005} (\bibinfo {year} {2016})}\BibitemShut {NoStop}%
\bibitem [{\citenamefont {Narevicius}\ \emph {et~al.}(2008)\citenamefont
  {Narevicius}, \citenamefont {Libson}, \citenamefont {Parthey}, \citenamefont
  {Chavez}, \citenamefont {Narevicius}, \citenamefont {Even},\ and\
  \citenamefont {Raizen}}]{Narevicius_2008}%
  \BibitemOpen
  \bibfield  {author} {\bibinfo {author} {\bibfnamefont {E.}~\bibnamefont
  {Narevicius}}, \bibinfo {author} {\bibfnamefont {A.}~\bibnamefont {Libson}},
  \bibinfo {author} {\bibfnamefont {C.~G.}\ \bibnamefont {Parthey}}, \bibinfo
  {author} {\bibfnamefont {I.}~\bibnamefont {Chavez}}, \bibinfo {author}
  {\bibfnamefont {J.}~\bibnamefont {Narevicius}}, \bibinfo {author}
  {\bibfnamefont {U.}~\bibnamefont {Even}}, \ and\ \bibinfo {author}
  {\bibfnamefont {M.~G.}\ \bibnamefont {Raizen}},\ }\href {\doibase
  10.1103/PhysRevLett.100.093003} {\bibfield  {journal} {\bibinfo  {journal}
  {Phys. Rev. Lett.}\ }\textbf {\bibinfo {volume} {100}},\ \bibinfo {pages}
  {093003} (\bibinfo {year} {2008})}\BibitemShut {NoStop}%
\bibitem [{\citenamefont {van~de Meerakker}\ \emph {et~al.}(2006)\citenamefont
  {van~de Meerakker}, \citenamefont {Vanhaecke},\ and\ \citenamefont
  {Meijer}}]{Meerakker_2006}%
  \BibitemOpen
  \bibfield  {author} {\bibinfo {author} {\bibfnamefont {S.~Y.}\ \bibnamefont
  {van~de Meerakker}}, \bibinfo {author} {\bibfnamefont {N.}~\bibnamefont
  {Vanhaecke}}, \ and\ \bibinfo {author} {\bibfnamefont {G.}~\bibnamefont
  {Meijer}},\ }\href {\doibase 10.1146/annurev.physchem.55.091602.094337}
  {\bibfield  {journal} {\bibinfo  {journal} {Ann. Rev. Phys. Chem.}\ }\textbf
  {\bibinfo {volume} {57}},\ \bibinfo {pages} {159} (\bibinfo {year}
  {2006})}\BibitemShut {NoStop}%
\bibitem [{\citenamefont {Zhu}\ \emph {et~al.}(1991)\citenamefont {Zhu},
  \citenamefont {Oates},\ and\ \citenamefont {Hall}}]{Zhu_1991}%
  \BibitemOpen
  \bibfield  {author} {\bibinfo {author} {\bibfnamefont {M.}~\bibnamefont
  {Zhu}}, \bibinfo {author} {\bibfnamefont {C.~W.}\ \bibnamefont {Oates}}, \
  and\ \bibinfo {author} {\bibfnamefont {J.~L.}\ \bibnamefont {Hall}},\ }\href
  {\doibase 10.1103/PhysRevLett.67.46} {\bibfield  {journal} {\bibinfo
  {journal} {Phys. Rev. Lett.}\ }\textbf {\bibinfo {volume} {67}},\ \bibinfo
  {pages} {46} (\bibinfo {year} {1991})}\BibitemShut {NoStop}%
\bibitem [{\citenamefont {Ertmer}\ \emph {et~al.}(1985)\citenamefont {Ertmer},
  \citenamefont {Blatt}, \citenamefont {Hall},\ and\ \citenamefont
  {Zhu}}]{Ertmer_1985}%
  \BibitemOpen
  \bibfield  {author} {\bibinfo {author} {\bibfnamefont {W.}~\bibnamefont
  {Ertmer}}, \bibinfo {author} {\bibfnamefont {R.}~\bibnamefont {Blatt}},
  \bibinfo {author} {\bibfnamefont {J.~L.}\ \bibnamefont {Hall}}, \ and\
  \bibinfo {author} {\bibfnamefont {M.}~\bibnamefont {Zhu}},\ }\href {\doibase
  10.1103/PhysRevLett.54.996} {\bibfield  {journal} {\bibinfo  {journal} {Phys.
  Rev. Lett.}\ }\textbf {\bibinfo {volume} {54}},\ \bibinfo {pages} {996}
  (\bibinfo {year} {1985})}\BibitemShut {NoStop}%
\bibitem [{\citenamefont {Petzold}\ \emph
  {et~al.}(2018{\natexlab{a}})\citenamefont {Petzold}, \citenamefont {Kaebert},
  \citenamefont {Gersema}, \citenamefont {Siercke},\ and\ \citenamefont
  {Ospelkaus}}]{Petzold_2018}%
  \BibitemOpen
  \bibfield  {author} {\bibinfo {author} {\bibfnamefont {M.}~\bibnamefont
  {Petzold}}, \bibinfo {author} {\bibfnamefont {P.}~\bibnamefont {Kaebert}},
  \bibinfo {author} {\bibfnamefont {P.}~\bibnamefont {Gersema}}, \bibinfo
  {author} {\bibfnamefont {M.}~\bibnamefont {Siercke}}, \ and\ \bibinfo
  {author} {\bibfnamefont {S.}~\bibnamefont {Ospelkaus}},\ }\href {\doibase
  10.1088/1367-2630/aab9f5} {\bibfield  {journal} {\bibinfo  {journal} {New J.
  Phys}\ }\textbf {\bibinfo {volume} {20}},\ \bibinfo {pages} {042001}
  (\bibinfo {year} {2018}{\natexlab{a}})}\BibitemShut {NoStop}%
\bibitem [{\citenamefont {Petzold}\ \emph
  {et~al.}(2018{\natexlab{b}})\citenamefont {Petzold}, \citenamefont {Kaebert},
  \citenamefont {Gersema}, \citenamefont {Poll}, \citenamefont {Reinhardt},
  \citenamefont {Siercke},\ and\ \citenamefont {Ospelkaus}}]{Petzold_2018_pra}%
  \BibitemOpen
  \bibfield  {author} {\bibinfo {author} {\bibfnamefont {M.}~\bibnamefont
  {Petzold}}, \bibinfo {author} {\bibfnamefont {P.}~\bibnamefont {Kaebert}},
  \bibinfo {author} {\bibfnamefont {P.}~\bibnamefont {Gersema}}, \bibinfo
  {author} {\bibfnamefont {T.}~\bibnamefont {Poll}}, \bibinfo {author}
  {\bibfnamefont {N.}~\bibnamefont {Reinhardt}}, \bibinfo {author}
  {\bibfnamefont {M.}~\bibnamefont {Siercke}}, \ and\ \bibinfo {author}
  {\bibfnamefont {S.}~\bibnamefont {Ospelkaus}},\ }\href {\doibase
  10.1103/PhysRevA.98.063408} {\bibfield  {journal} {\bibinfo  {journal} {Phys.
  Rev. A}\ }\textbf {\bibinfo {volume} {98}},\ \bibinfo {pages} {063408}
  (\bibinfo {year} {2018}{\natexlab{b}})}\BibitemShut {NoStop}%
\bibitem [{\citenamefont {Chen}\ \emph {et~al.}(2016)\citenamefont {Chen},
  \citenamefont {Bu},\ and\ \citenamefont {Yan}}]{Chen_2016}%
  \BibitemOpen
  \bibfield  {author} {\bibinfo {author} {\bibfnamefont {T.}~\bibnamefont
  {Chen}}, \bibinfo {author} {\bibfnamefont {W.}~\bibnamefont {Bu}}, \ and\
  \bibinfo {author} {\bibfnamefont {B.}~\bibnamefont {Yan}},\ }\href {\doibase
  10.1103/PhysRevA.94.063415} {\bibfield  {journal} {\bibinfo  {journal} {Phys.
  Rev. A}\ }\textbf {\bibinfo {volume} {94}},\ \bibinfo {pages} {063415}
  (\bibinfo {year} {2016})}\BibitemShut {NoStop}%
\bibitem [{\citenamefont {Bu}\ \emph {et~al.}(2017)\citenamefont {Bu},
  \citenamefont {Chen}, \citenamefont {Lv},\ and\ \citenamefont
  {Yan}}]{Bu_2017}%
  \BibitemOpen
  \bibfield  {author} {\bibinfo {author} {\bibfnamefont {W.}~\bibnamefont
  {Bu}}, \bibinfo {author} {\bibfnamefont {T.}~\bibnamefont {Chen}}, \bibinfo
  {author} {\bibfnamefont {G.}~\bibnamefont {Lv}}, \ and\ \bibinfo {author}
  {\bibfnamefont {B.}~\bibnamefont {Yan}},\ }\href {\doibase
  10.1103/PhysRevA.95.032701} {\bibfield  {journal} {\bibinfo  {journal} {Phys.
  Rev. A}\ }\textbf {\bibinfo {volume} {95}},\ \bibinfo {pages} {032701}
  (\bibinfo {year} {2017})}\BibitemShut {NoStop}%
\bibitem [{\citenamefont {Chen}\ \emph {et~al.}(2017)\citenamefont {Chen},
  \citenamefont {Bu},\ and\ \citenamefont {Yan}}]{Chen2017}%
  \BibitemOpen
  \bibfield  {author} {\bibinfo {author} {\bibfnamefont {T.}~\bibnamefont
  {Chen}}, \bibinfo {author} {\bibfnamefont {W.}~\bibnamefont {Bu}}, \ and\
  \bibinfo {author} {\bibfnamefont {B.}~\bibnamefont {Yan}},\ }\href {\doibase
  10.1103/PhysRevA.96.053401} {\bibfield  {journal} {\bibinfo  {journal} {Phys.
  Rev. A}\ }\textbf {\bibinfo {volume} {96}},\ \bibinfo {pages} {053401}
  (\bibinfo {year} {2017})}\BibitemShut {NoStop}%
\bibitem [{\citenamefont {Kozyryev}\ \emph {et~al.}(2015)\citenamefont
  {Kozyryev}, \citenamefont {Baum}, \citenamefont {Matsuda}, \citenamefont
  {Olson}, \citenamefont {Hemmerling},\ and\ \citenamefont
  {Doyle}}]{Kozyryev_2015}%
  \BibitemOpen
  \bibfield  {author} {\bibinfo {author} {\bibfnamefont {I.}~\bibnamefont
  {Kozyryev}}, \bibinfo {author} {\bibfnamefont {L.}~\bibnamefont {Baum}},
  \bibinfo {author} {\bibfnamefont {K.}~\bibnamefont {Matsuda}}, \bibinfo
  {author} {\bibfnamefont {P.}~\bibnamefont {Olson}}, \bibinfo {author}
  {\bibfnamefont {B.}~\bibnamefont {Hemmerling}}, \ and\ \bibinfo {author}
  {\bibfnamefont {J.~M.}\ \bibnamefont {Doyle}},\ }\href {\doibase
  10.1088/1367-2630/17/4/045003} {\bibfield  {journal} {\bibinfo  {journal}
  {New J. Phys}\ }\textbf {\bibinfo {volume} {17}},\ \bibinfo {pages} {045003}
  (\bibinfo {year} {2015})}\BibitemShut {NoStop}%
\bibitem [{\citenamefont {Kozyryev}\ \emph {et~al.}(2016)\citenamefont
  {Kozyryev}, \citenamefont {Baum}, \citenamefont {Matsuda}, \citenamefont
  {Hemmerling},\ and\ \citenamefont {Doyle}}]{Kozyryev_2016}%
  \BibitemOpen
  \bibfield  {author} {\bibinfo {author} {\bibfnamefont {I.}~\bibnamefont
  {Kozyryev}}, \bibinfo {author} {\bibfnamefont {L.}~\bibnamefont {Baum}},
  \bibinfo {author} {\bibfnamefont {K.}~\bibnamefont {Matsuda}}, \bibinfo
  {author} {\bibfnamefont {B.}~\bibnamefont {Hemmerling}}, \ and\ \bibinfo
  {author} {\bibfnamefont {J.~M.}\ \bibnamefont {Doyle}},\ }\href {\doibase
  10.1088/0953-4075/49/13/134002} {\bibfield  {journal} {\bibinfo  {journal}
  {J. Phys. B}\ }\textbf {\bibinfo {volume} {49}},\ \bibinfo {pages} {134002}
  (\bibinfo {year} {2016})}\BibitemShut {NoStop}%
\bibitem [{\citenamefont {Phillips}\ and\ \citenamefont
  {Metcalf}(1982)}]{Phillips_1982}%
  \BibitemOpen
  \bibfield  {author} {\bibinfo {author} {\bibfnamefont {W.~D.}\ \bibnamefont
  {Phillips}}\ and\ \bibinfo {author} {\bibfnamefont {H.}~\bibnamefont
  {Metcalf}},\ }\href {\doibase 10.1103/PhysRevLett.48.596} {\bibfield
  {journal} {\bibinfo  {journal} {Phys. Rev. Lett.}\ }\textbf {\bibinfo
  {volume} {48}},\ \bibinfo {pages} {596} (\bibinfo {year} {1982})}\BibitemShut
  {NoStop}%
\bibitem [{\citenamefont {Berkeland}\ and\ \citenamefont
  {Boshier}(2002)}]{Berkeland_2002}%
  \BibitemOpen
  \bibfield  {author} {\bibinfo {author} {\bibfnamefont {D.~J.}\ \bibnamefont
  {Berkeland}}\ and\ \bibinfo {author} {\bibfnamefont {M.~G.}\ \bibnamefont
  {Boshier}},\ }\href {\doibase 10.1103/PhysRevA.65.033413} {\bibfield
  {journal} {\bibinfo  {journal} {Phys. Rev. A}\ }\textbf {\bibinfo {volume}
  {65}},\ \bibinfo {pages} {033413} (\bibinfo {year} {2002})}\BibitemShut
  {NoStop}%
\bibitem [{\citenamefont {Shuman}\ \emph {et~al.}(2010)\citenamefont {Shuman},
  \citenamefont {Barry},\ and\ \citenamefont {DeMille}}]{Shuman_2010}%
  \BibitemOpen
  \bibfield  {author} {\bibinfo {author} {\bibfnamefont {E.~S.}\ \bibnamefont
  {Shuman}}, \bibinfo {author} {\bibfnamefont {J.~F.}\ \bibnamefont {Barry}}, \
  and\ \bibinfo {author} {\bibfnamefont {D.}~\bibnamefont {DeMille}},\ }\href
  {\doibase 10.1038/nature09443} {\bibfield  {journal} {\bibinfo  {journal}
  {Nature}\ }\textbf {\bibinfo {volume} {467}},\ \bibinfo {pages} {820}
  (\bibinfo {year} {2010})}\BibitemShut {NoStop}%
\bibitem [{\citenamefont {Hummon}\ \emph {et~al.}(2013)\citenamefont {Hummon},
  \citenamefont {Yeo}, \citenamefont {Stuhl}, \citenamefont {Collopy},
  \citenamefont {Xia},\ and\ \citenamefont {Ye}}]{Hummon_2013}%
  \BibitemOpen
  \bibfield  {author} {\bibinfo {author} {\bibfnamefont {M.~T.}\ \bibnamefont
  {Hummon}}, \bibinfo {author} {\bibfnamefont {M.}~\bibnamefont {Yeo}},
  \bibinfo {author} {\bibfnamefont {B.~K.}\ \bibnamefont {Stuhl}}, \bibinfo
  {author} {\bibfnamefont {A.~L.}\ \bibnamefont {Collopy}}, \bibinfo {author}
  {\bibfnamefont {Y.}~\bibnamefont {Xia}}, \ and\ \bibinfo {author}
  {\bibfnamefont {J.}~\bibnamefont {Ye}},\ }\href {\doibase
  10.1103/PhysRevLett.110.143001} {\bibfield  {journal} {\bibinfo  {journal}
  {Phys. Rev. Lett.}\ }\textbf {\bibinfo {volume} {110}},\ \bibinfo {pages}
  {143001} (\bibinfo {year} {2013})}\BibitemShut {NoStop}%
\bibitem [{\citenamefont {Truppe}\ \emph {et~al.}(2018)\citenamefont {Truppe},
  \citenamefont {Hambach}, \citenamefont {Skoff}, \citenamefont {Bulleid},
  \citenamefont {Bumby}, \citenamefont {Hendricks}, \citenamefont {Hinds},
  \citenamefont {Sauer},\ and\ \citenamefont {Tarbutt}}]{Truppe_2018}%
  \BibitemOpen
  \bibfield  {author} {\bibinfo {author} {\bibfnamefont {S.}~\bibnamefont
  {Truppe}}, \bibinfo {author} {\bibfnamefont {M.}~\bibnamefont {Hambach}},
  \bibinfo {author} {\bibfnamefont {S.~M.}\ \bibnamefont {Skoff}}, \bibinfo
  {author} {\bibfnamefont {N.~E.}\ \bibnamefont {Bulleid}}, \bibinfo {author}
  {\bibfnamefont {J.~S.}\ \bibnamefont {Bumby}}, \bibinfo {author}
  {\bibfnamefont {R.~J.}\ \bibnamefont {Hendricks}}, \bibinfo {author}
  {\bibfnamefont {E.~A.}\ \bibnamefont {Hinds}}, \bibinfo {author}
  {\bibfnamefont {B.~E.}\ \bibnamefont {Sauer}}, \ and\ \bibinfo {author}
  {\bibfnamefont {M.~R.}\ \bibnamefont {Tarbutt}},\ }\href {\doibase
  10.1080/09500340.2017.1384516} {\bibfield  {journal} {\bibinfo  {journal} {J.
  Mod. Opt.}\ }\textbf {\bibinfo {volume} {65}},\ \bibinfo {pages} {648}
  (\bibinfo {year} {2018})}\BibitemShut {NoStop}%
\bibitem [{\citenamefont {Scully}\ and\ \citenamefont
  {Zubairy}(1997)}]{Scully_1997}%
  \BibitemOpen
  \bibfield  {author} {\bibinfo {author} {\bibfnamefont {M.~O.}\ \bibnamefont
  {Scully}}\ and\ \bibinfo {author} {\bibfnamefont {M.~S.}\ \bibnamefont
  {Zubairy}},\ }\href@noop {} {\emph {\bibinfo {title} {Quantum Optics}}}\
  (\bibinfo  {publisher} {Cambridge University Press},\ \bibinfo {year}
  {1997})\BibitemShut {NoStop}%
\bibitem [{\citenamefont {Barry}\ \emph {et~al.}(2011)\citenamefont {Barry},
  \citenamefont {Shuman},\ and\ \citenamefont {DeMille}}]{Barry_2011}%
  \BibitemOpen
  \bibfield  {author} {\bibinfo {author} {\bibfnamefont {J.~F.}\ \bibnamefont
  {Barry}}, \bibinfo {author} {\bibfnamefont {E.~S.}\ \bibnamefont {Shuman}}, \
  and\ \bibinfo {author} {\bibfnamefont {D.}~\bibnamefont {DeMille}},\ }\href
  {\doibase 10.1039/C1CP20335E} {\bibfield  {journal} {\bibinfo  {journal}
  {Phys. Chem. Chem. Phys.}\ }\textbf {\bibinfo {volume} {13}},\ \bibinfo
  {pages} {18936} (\bibinfo {year} {2011})}\BibitemShut {NoStop}%
\bibitem [{\citenamefont {Fletcher}\ \emph {et~al.}(1993)\citenamefont
  {Fletcher}, \citenamefont {Jung}, \citenamefont {Scurlock},\ and\
  \citenamefont {Steimle}}]{Fletcher_1993}%
  \BibitemOpen
  \bibfield  {author} {\bibinfo {author} {\bibfnamefont {D.~A.}\ \bibnamefont
  {Fletcher}}, \bibinfo {author} {\bibfnamefont {K.~Y.}\ \bibnamefont {Jung}},
  \bibinfo {author} {\bibfnamefont {C.~T.}\ \bibnamefont {Scurlock}}, \ and\
  \bibinfo {author} {\bibfnamefont {T.~C.}\ \bibnamefont {Steimle}},\ }\href
  {\doibase 10.1063/1.464218} {\bibfield  {journal} {\bibinfo  {journal} {J.
  Chem. Phys.}\ }\textbf {\bibinfo {volume} {98}},\ \bibinfo {pages} {1837}
  (\bibinfo {year} {1993})}\BibitemShut {NoStop}%
\bibitem [{\citenamefont {Presunka}\ and\ \citenamefont
  {Coxon}(1995)}]{Presunka_1995}%
  \BibitemOpen
  \bibfield  {author} {\bibinfo {author} {\bibfnamefont {P.~I.}\ \bibnamefont
  {Presunka}}\ and\ \bibinfo {author} {\bibfnamefont {J.~A.}\ \bibnamefont
  {Coxon}},\ }\href {\doibase https://doi.org/10.1016/0301-0104(94)00330-D}
  {\bibfield  {journal} {\bibinfo  {journal} {Chem. Phys}\ }\textbf {\bibinfo
  {volume} {190}},\ \bibinfo {pages} {97 } (\bibinfo {year}
  {1995})}\BibitemShut {NoStop}%
\bibitem [{\citenamefont {Kozyryev}\ \emph {et~al.}(2017)\citenamefont
  {Kozyryev}, \citenamefont {Baum}, \citenamefont {Matsuda}, \citenamefont
  {Augenbraun}, \citenamefont {Anderegg}, \citenamefont {Sedlack},\ and\
  \citenamefont {Doyle}}]{Kozyryev_2017}%
  \BibitemOpen
  \bibfield  {author} {\bibinfo {author} {\bibfnamefont {I.}~\bibnamefont
  {Kozyryev}}, \bibinfo {author} {\bibfnamefont {L.}~\bibnamefont {Baum}},
  \bibinfo {author} {\bibfnamefont {K.}~\bibnamefont {Matsuda}}, \bibinfo
  {author} {\bibfnamefont {B.~L.}\ \bibnamefont {Augenbraun}}, \bibinfo
  {author} {\bibfnamefont {L.}~\bibnamefont {Anderegg}}, \bibinfo {author}
  {\bibfnamefont {A.~P.}\ \bibnamefont {Sedlack}}, \ and\ \bibinfo {author}
  {\bibfnamefont {J.~M.}\ \bibnamefont {Doyle}},\ }\href {\doibase
  10.1103/PhysRevLett.118.173201} {\bibfield  {journal} {\bibinfo  {journal}
  {Phys. Rev. Lett.}\ }\textbf {\bibinfo {volume} {118}},\ \bibinfo {pages}
  {173201} (\bibinfo {year} {2017})}\BibitemShut {NoStop}%
\end{thebibliography}%


%
\end{document}